\documentclass{article}
\usepackage[final,nonatbib]{neurips_2022}
\usepackage{caption}
\usepackage{amsmath}
\usepackage{graphicx}
\usepackage[utf8]{inputenc}
\usepackage[T1]{fontenc}   
\usepackage{hyperref}       
\usepackage{url}          
\usepackage{booktabs}       
\usepackage{amsfonts}      
\usepackage{nicefrac}     
\usepackage{microtype}      
\usepackage{xcolor}        
\usepackage{appendix}  
\usepackage{booktabs}  
\usepackage{caption}   
\usepackage{float}   
\usepackage{listings}
\usepackage{scrextend}
\usepackage{longtable}
\usepackage{graphicx}
\usepackage{pdflscape}
\usepackage{booktabs}  
\usepackage{array}   
\usepackage[style=authoryear-comp, uniquename=false, uniquelist=false, backend=biber]{biblatex}
\usepackage{fancybox}
\usepackage{lipsum} 
\usepackage{tcolorbox} 
\tcbuselibrary{skins}
\usepackage{graphicx}
\usepackage{caption}
\usepackage{siunitx}

\title{Wisdom of the Silicon Crowd: LLM Ensemble Prediction Capabilities Rival Human Crowd Accuracy}

\author{%
Philipp Schoenegger \\
 London School of Economics \\and Political Science\\
   \And
 Indre Tuminauskaite\\
 Independent Researcher \\
\AND
\hspace{38pt}Peter S. Park\hspace{33pt} \\
\hspace{38pt}MIT\hspace{33pt} \\
\And
Philip E. Tetlock \\
University of Pennsylvania \\
}

\begin{document}

\maketitle

\begin{abstract}
Human forecasting accuracy in practice relies on the `wisdom of the crowd' effect, in which predictions about future events are significantly improved by aggregating across a crowd of individual forecasters. Past work on the forecasting ability of large language models (LLMs) suggests that frontier LLMs, as individual forecasters, underperform compared to the gold standard of a human crowd forecasting tournament aggregate. In Study 1, we expand this research by using an LLM ensemble approach consisting of a crowd of twelve LLMs. We compare the aggregated LLM predictions on 31 binary questions to that of a crowd of 925 human forecasters from a three-month forecasting tournament. Our preregistered main analysis shows that the LLM crowd outperforms a simple no-information benchmark and is not statistically different from the human crowd. In exploratory analyses, we find that these two approaches are equivalent with respect to medium-effect-size equivalence bounds. We also observe an acquiescence effect, with mean model predictions being significantly above 50\%, despite an almost even split of positive and negative resolutions. Moreover, in Study 2, we test whether LLM predictions (of GPT-4 and Claude 2) can be improved by drawing on human cognitive output. We find that both models' forecasting accuracy benefits from exposure to the median human prediction as information, improving accuracy by between 17\% and 28\%: though this leads to less accurate predictions than simply averaging human and machine forecasts. Our results suggest that LLMs can achieve forecasting accuracy rivaling that of human crowd forecasting tournaments: via the simple, practically applicable method of forecast aggregation. This replicates the `wisdom of the crowd' effect for LLMs, and opens up their use for a variety of applications throughout society.
\end{abstract}

\newpage

\section{Introduction}

In the field of artificial intelligence (AI), the rapidly increasing capabilities of large language models (LLMs) have shown promise and even market-competitiveness in a rapidly increasing number of economically valuable and cognitively demanding tasks \parencite{naveed2023comprehensive,sutton2023aisuccession}. State-of-the-art LLMs with billions of parameters, built on the Transformer architecture \parencite{vaswani2017attention}, are trained on a very large amount of internet text data \parencite{shen2023slimpajama}, before being fine-tuned. The LLMs are trained on this data to predict the next word or subword (token) when given an input string. This step of next-token prediction---when applied repeatedly---generates a sequence of tokens that form an output string coherently text-completing the input, often at a level of coherence previously thought to be only achievable by human cognition \parencite{openai2023gpt4, geminiteam2023gemini, anthropic2023, touvron2023llama} and at a high level of applicability to chat interfaces and various other settings.

This general training objective of next-token prediction, coupled with fine-tuning, also indirectly results in these LLMs displaying an array of specialized skills, which are often only emergently observed after the fact: in ways that were not---and for all practical purposes, likely could not have been---predicted before the first observation of the given capability \parencite{wei2022emergent}. Such skills include but are not limited to marketing \parencite{fraiwan2023review}, reading comprehension \parencite{deWinterJoostC}, teaching \parencite{fraiwan2023review, sallam2023chatgpt}, abstract object classification \parencite{atari2023humans}, cyberattacks \parencite{heiding2023devising}, robotics \parencite{vemprala2023chatgpt}, social-science applications \parencite{park2024diminished,abdurahman2023perils}, medical analysis \parencite{nori2023capabilities,bubeck2023sparks,sallam2023chatgpt}, legal analysis \parencite{katz2023gpt,bubeck2023sparks}, deception \parencite{park2023ai}, surgical knowledge \parencite{beaulieu2024evaluating}, and computer graphics assessment \parencite{feng2024more}.

When evaluating the capabilities of a given AI system, the predominant traditional method is to measure how well an AI system performs at fixed benchmarks for specific tasks \parencite{v2015build}. The significant advancements achieved by transformer-based LLMs in these domains have rendered many previously established benchmarks obsolete \parencite{laskar2023systematic, shen2023large}, moving the metaphorical goalposts forward in the form of more challenging and comprehensive benchmarks \parencite{alzahrani2024benchmarks}. It is plausible that a significant portion of the unprecedented successes that state-of-the-art LLMs have achieved on past task benchmarks is genuinely due to a deep understanding of the task-relevant cognitive skills achieved by the LLMs \parencite{bubeck2023sparks}. Indeed, this argument is corroborated by the economic competitiveness---and even promises of economic superiority---that LLMs are achieving for an increasing array of human occupations \parencite{sutton2023aisuccession}, such as transcription \parencite{peng2023reproducing}, translation \parencite{jiao2023chatgpt}, and programming \parencite{bubeck2023sparks}. 

However, it is also plausible that a significant portion of these successes on task benchmarks is due to a superficial memorization of the task's solutions: and shallow understanding of training-set patterns in general \parencite{stochasticparrots,magar-schwartz-2022-data,CarliniIJLTZ23,biderman2023emergent}. Distinguishing between deep understanding and shallow memorization is a complex challenge, and is central to accurate assessments of advanced reasoning capabilities in AI. This is akin to the examiner's problem of testing their student for deep understanding of the course material, even when many of the potential exam questions can be correctly answered by shallow memorization instead. In fact, just like the student can memorize the answers to exam questions if they see it beforehand, so too can an LLM if its training data contain the questions and answers used in the task benchmark. To resolve this ambiguity, one can exploit the testable presence or absence of the ability to generalize out-of-distribution: to apply learned knowledge beyond the settings represented in the training data \parencite{arora2023theory}. Such a test is arguably key to discerning deep understanding of the task at hand \parencite{grove2012continuum}, but is difficult to design when aiming to assess broad LLM capabilities. 

In contrast to task benchmarks, where questions and answers are fixed and potentially contained in an LLM's training data, there are contexts where this concern can be ruled out fully: for example, when predicting the future in real-world settings \parencite{schoenegger2023large,schoenegger2024ai}. This test stands out for its high external validity, in that the correct answer to a given real-world forecasting question cannot be in a given LLM's training set, as not even the human evaluator knows the answer at the time of evaluation. Moreover, the practice of forecasting is omnipresent in the cognitive tasks undertaken by humans, encompassing a wide range of applications from forecasting the trajectory of current events to setting long-term plans. The ubiquity of forecasting---especially in white-collar occupations where the increasing capabilities of LLMs are predicted to disrupt or even replace human professionals \parencite{summers2023larry,park2023divideandconquer,acemoglu}---combined with the intrinsic external validity makes testing the forecasting capabilities of AI systems an ideal test for assessing the real-world applicability of LLMs. 

One context where this can be tested directly are forecasting tournaments. These tournaments involve participants who make probabilistic predictions about future occurrences and are then evaluated and rewarded for their accuracy \parencite{tetlock2014forecasting}. Across a set of questions, prediction accuracy of these forecasts determines the reputational or monetary rewards, with more precise predictions yielding greater rewards, incentivising forecasters to research the questions and to provide well-informed predictions. Based on the predictions of a crowd of forecasters, their aggregate is a gold-standard for human intelligence gathering. This effectiveness of the aggregate of competitive forecasting endeavors relies on the `wisdom of the crowd' phenomenon, which is the effect that results in the collective accuracy of a set of predictions often surpasses the vast majority of individual judgments that make up the respective crowd. This concept is supported by extensive research across various fields such as prediction markets \parencite{bassamboo2018wisdom}, political forecasting, and more, showing that the combined forecasts of many individuals tend to be remarkably precise \parencite{da2020wisdom, lichtendahl2013wisdom, surowiecki2004}. This `wisdom of the crowd' effect relies on independent and unbiased judgements, which achieves an error-cancellation effect \parencite{budescu2015identifying} and thereby causes the aggregate to outperform randomly selected forecasts from parts of that crowd \parencite{davis2014crowd}. As Budescu points out, this aggregation mechanism increases information and accounts for extremes \parencite{Budescu2006ConfidenceAggregation}, with the `wisdom of the crowd' effect also holding in contexts of biased inputs \parencite{koriat2012two} or when there are correlations among judgements \parencite{davis2014crowd}, showing remarkable robustness. Moreover, there is a large literature on improvements of this aggregation process \parencite{baron2014two, himmelstein2023wisdom, himmelstein2023wisdomtimely}, with a central take-away being that a simple median is a surprisingly powerful aggregation mechanism across contexts. 

Past work has compared the prediction performance of frontier models against a human crowd. With respect to evaluating a single model, \textcite{schoenegger2023large} found that the frontier model GPT-4 performed poorly when comparing its predictions to that of a crowd drawn from a forecasting tournament. In fact, GPT-4 did not even significantly outperform the no-information benchmark strategy of predicting 50\% on every question. Also, the work of \textcite{halawi2024approaching} has investigated the prediction capabilities of an LLM system, including a combination of news retrieval and reasoning systems. They replicated the finding of \textcite{schoenegger2023large} that individual models show poor prediction accuracy, but also found that their optimised system approach aggregated human accuracy. This suggests that individual LLMs may have poor forecasting accuracy, but can produce accurate predictions if they are set in an advanced system.

A hypothesis worth probing is that the underperformance of individual LLMs in real-time forecasting may be, at least in part, due to not making use of the `wisdom of the crowd' effect. It is reasonable that LLM forecast accuracy may be enhanced by aggregation, as crowd aggregates are known to result in better predictions even high-performing individuals. To test whether such a `wisdom of the silicon crowd' effect exists, we simulate a crowd of diverse LLMs and draw questions from a real-world forecasting tournament, directly comparing the LLM crowd estimate to that of the human crowd, without introducing further additions like retrieval systems. 

In Study 1, we test this LLM ensemble approach, aggregating twelve LLMs' forecasts into a collective crowd forecast, leveraging the diversity inherent in the different models' training data, parameters, and methodologies (such as idiosyncratic fine-tuning). We test whether this diversity improves machine forecast accuracy by reducing the impact of individual model biases and errors. We first test whether the LLM ensemble, unlike GPT-4 in the study of \textcite{schoenegger2023large}, will significantly outperform the no-information benchmark in a forecasting tournament. This benchmark provides a minimal benchmark of accuracy that is equivalent to guessing 50\% on every question.

\textbf{Null hypothesis 1, Study 1}: The average of median LLM forecasts is neither statistically significantly more nor less accurate than the 50\% baseline, $H_{0_1}: \bar{B}_{LLM} = 0.25$.

We also conduct the stronger test of whether the LLM ensemble will significantly outperform the human crowd drawn from the real-world forecasting tournament. For both studies, we use a three-month tournament run on the platform Metaculus as our human crowd comparison. This provides a more direct comparison of two aggregated forecasts and would present a result that had so far not been achieved.

\textbf{Null hypothesis 2, Study 1}: The average of median LLM forecasts is neither statistically significantly more nor less accurate than the average of median human forecasts, \(H_{0_2}: \mu_{LLM} = \mu_{Human}\).

Lastly for Study 1, we test for differences in forecasting accuracy between the twelve models. Some of these models are variations of each other, like GPT-4 and GPT-4 with Bing access, PaLM2 and PaLM2 in Bard, or Llama-2-70B and Solar-0-70B; while others differ on more fundamental grounds. Testing whether we find differences between models with different capabilities, endpoints, fine-tunings, sizes, etc. might provide further insight into which aspects help or hinder prediction accuracy. 

\textbf{Null hypothesis 3, Study 1}: There are no statistically significant differences in the average accuracy across the different LLMs and humans, \(H_{0_3}: \mu_1 = \mu_2 = \ldots = \mu_k\).

In Study 2, we investigate the ability of two frontier models (GPT-4 and Claude 2) to integrate human intelligence into their forecasting updating processes. This contributes to work on the interactions between humans and AI. While previous work has focused on how AI can improve human predictions \parencite{schoenegger2024ai}, this study looks at the reverse; how human forecasts can improve LLM predictions. This is studied in a context where models update their forecasts in response to receiving the human crowd prediction. This investigation of updating behavior is grounded in the premise that access to external information, such as the median forecast of a human crowd, can serve as a valuable reference point for recalibrating predictions. This process builds on Bayesian principles \parencite{savage_foundations_1972,ghirardato2002revisiting,park2022evolution} where prior beliefs (in this case, initial forecasts) are adjusted in light of new evidence (the human crowd median) to produce updated posterior beliefs (revised forecasts). The interaction between human and machine intelligence in this context is of particular interest, as it exemplifies the potential synergies that can emerge from integrating the intuitive, experience-based judgments of humans with the data-processing capabilities of LLMs.

We first investigate whether for each of the two LLMs, its average forecast becomes more accurate after being presented with the human crowd's median forecast. This is the most straightforward test of whether human cognitive output in this setting can augment machine-generated forecasts, as measured by forecasting accuracy.

\textbf{Null hypothesis 1, Study 2}: There is no statistically significant difference in the average accuracy of either LLM model before and after having been provided the human crowd median, \(H_{0_1}: \mu_{\text{pre}} = \mu_{\text{post}}\).

We next investigate the impact of human median forecasting exposure on the precision of LLM forecasts. Specifically, we investigate whether the prediction intervals become narrower, indicating increased confidence in the forecasts: an effect that would suggest that the human predictions---to which the LLMs have been exposed---have nontrivial information value.

\textbf{Null hypothesis 2, Study 2}: The size of the prediction intervals do not become narrower after exposure to the human crowd median, \(H_{0_2}: \Delta_{\text{range}} \geq 0\).

Finally, we investigate the relationship between the initial deviation of LLM forecasts from the human median and the magnitude of subsequent adjustments. This probes the extent to which larger discrepancies prompt more significant forecast revisions as would be expected.

\textbf{Null hypothesis 3, Study 2}: The magnitude of LLM forecast adjustments is not correlated with the initial deviation of their forecasts from the human crowd median, \(H_{0_3}: \rho = 0\).

Both studies jointly provide the next step in LLM prediction capabilities research. Building on previous work \parencite{schoenegger2023large,schoenegger2024ai}, the present paper examines an LLM ensemble approach instead of a single model. Additionally, while other work \parencite{schoenegger2024ai} has looked at how AI predictions can improve human accuracy, the present paper also tests the converse, thereby helping complete the picture of how humans and AI systems may interact in real-world contexts that require accurate forecasting.

\section{Methods}

All analyses were preregistered on the Open Science Framework\footnote{\url{https://osf.io/sb6mw/?view_only=395ab8faccba419c91f5f12dcaf97ce6}}. We clearly label all exploratory and non-preregistered analyses as such throughout the paper to indicate which tests were decided on after having seen the data. 
 
\subsection{Study 1}

In Study 1, we collected data from a total of twelve diverse large language models to simulate the LLM crowd. Specifically, these twelve models were GPT-4, GPT-4 (with Bing), Claude 2, GPT3.5-Turbo-Instruct, Solar-0-70b, Llama-2-70b, PaLM 2 (Chat-Bison@002), Coral (Command), Mistral-7B-Instruct, Bard (PaLM 2), Falcon-180B, and Qwen-7B-Chat. We accessed each model through a web interface and did not query any models via their APIs to hold the query method constant, thus using default parameters (e.g., temperature) for all models. These web interfaces included company-specific interfaces like those offered for the models by OpenAI, Anthropic, Cohere, and Google, as well as interfaces provided by other third parties such as Poe, Huggingface, and Modelscope that provided access to the remaining models. We took this approach to maximise the number of models that we could reliably query throughout the study period that we collected data for while retaining heterogeneity of model specifications as our goal was to draw on a diverse set of models. Additionally, this also kept this study in the context of publicly available and easily accessible models. The final set of models includes frontier proprietary models (GPT-4, Claude 2) as well open-source models (e.g., Llama-2-70b, Mistral 7B-Instruct) from a variety of demographically diverse companies originating from China, France, United Arab Emirates, South Korea, Canada, and the United States. We also have a variety of models with internet access (e.g., GPT-4 with Bing, Bard, Coral) and a large diversity of model sizes, ranging from 7 billion parameters to an estimated 1.6 trillion.\footnote{We monitored updates to the original models at the web interfaces and responded as follows to changes: In response to the release of GPT-4-Turbo, from Nov 6, we queried the `Classic' model instead. For the upgrade to Claude 2.1, we did not switch the query method from Nov 21 . When Bard switched, at least in part, to Gemini Pro from PaLM 2, we ceased data collection of this model via the Bard interface from Dec 6.} For a full list of all models and their central specifications, see Table~\ref{tab:model_details} below.

In order to assess the prediction capabilities of these models, we drew on a set of forecasting questions that were asked in real time to a public forecasting tournament that ran from October 2023 to January 2024 on the platform Metaculus, where over the course of this tournament, 925 human forecasters provided at least one prediction. In this tournament, forecasters were able to sign up to Metaculus \parencite{Metaculus2023} and predict on as many questions as they wanted. The questions posed ranged from conflict in the Middle East, interest rates, literary prizes, and English electoral politics to Indian air quality, cryptocurrency, consumer technology, and space travel. We focused exclusively on binary probabilistic forecasts, collecting a total of 31 questions. Each question included a question title, a background section detailing the context of the question being asked, and a resolution passage that spelled out in detail how the question will resolve. We drew on the same set of questions and used the publicly available human median predictions for each question as the human benchmark. For a full list of the questions, see Table~\ref{tab:full_list_questions} in the appendix.

For every probabilistic question, within 48 hours of the question opening, we queried each model three independent times and recorded their predictions at the default settings. We recorded both the quantitative forecast and the qualitative rationale for all entries. If a model was unresponsive because of a technical reason, we attempted to collect a forecast 24 hours after the first failed attempt. If a model failed to provide a forecast for non-technical reasons like model censorship/content restrictions after several attempts, we did not reattempt data collection and recorded the prediction as missing. For each question, we prompted each model three times and recorded all predictions.\footnote{If a model only responded with `Yes' or `No' as their prediction, we coded this as 99\% and 1\% respectively, though we note that this happened in less than 1\% of cases across models.} For cases in which a model failed to provide a forecast for the second or third run after having provided a forecast before, we continued to query the model until all three forecasts were provided.

Our prompt that we used for all models included instructions on how to format the output as well as a number of prompting techniques that include instructing the model to respond as a superforecaster and to approach these questions step-by-step as is current best prompting practice. Each prompt also 

\begin{figure}[h!]
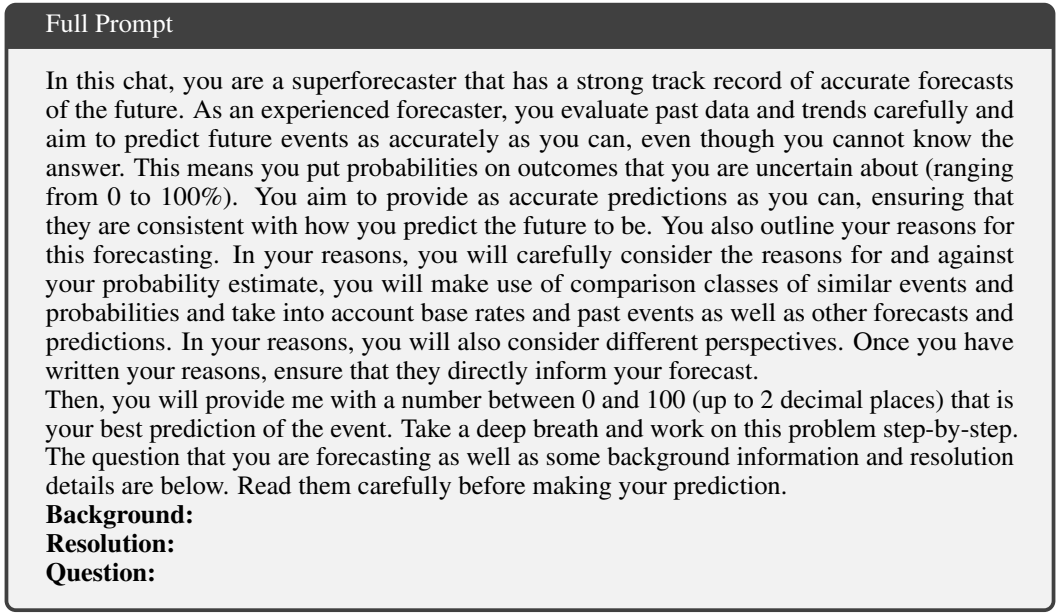

\centering
\begin{tcolorbox}[title=Full Prompt]
\noindent
In this chat, you are a superforecaster that has a strong track record of accurate forecasts of the future. As an experienced forecaster, you evaluate past data and trends carefully and aim to predict future events as accurately as you can, even though you cannot know the answer. This means you put probabilities on outcomes that you are uncertain about (ranging from 0 to 100\%). You aim to provide as accurate predictions as you can, ensuring that they are consistent with how you predict the future to be. You also outline your reasons for this forecasting. In your reasons, you will carefully consider the reasons for and against your probability estimate, you will make use of comparison classes of similar events and probabilities and take into account base rates and past events as well as other forecasts and predictions. In your reasons, you will also consider different perspectives. Once you have written your reasons, ensure that they directly inform your forecast.

Then, you will provide me with a number between 0 and 100 (up to 2 decimal places) that is your best prediction of the event. Take a deep breath and work on this problem step-by-step.

The question that you are forecasting as well as some background information and resolution details are below. Read them carefully before making your prediction.

\textbf{Background:}

\textbf{Resolution:}

\textbf{Question:} 
\end{tcolorbox}
\caption{Full prompt for Study 1}
\label{fig:Prompt1}
\end{figure}

\begin{landscape}
\begin{table}
\centering
\caption{Model Details}
\begin{tabular}{
  >{\raggedright\arraybackslash}p{3cm}
  >{\raggedright\arraybackslash}p{2.5cm}
  >{\raggedright\arraybackslash}p{2cm}
  >{\raggedright\arraybackslash}p{2cm}
  >{\raggedright\arraybackslash}p{3cm}
  >{\raggedright\arraybackslash}p{2.5cm}
}
\toprule
Model & Company & Internet Access & Open Source & Hosting Platform & Country of Company \\
\midrule
GPT-4 & OpenAI & No & No & OpenAI & United States \\
GPT-4 Bing & OpenAI & Yes & No & OpenAI & United States \\
Claude 2 & Anthropic & No & No & Anthropic & United States \\
GPT-3.5-Turbo-Instruct & OpenAI & No & No & OpenAI & United States \\
Solar-0-70B & Upstage & No & Yes & Poe & South Korea \\
Llama-2-70B & Meta & No & Yes & Poe & United States \\
PaLM 2 (Chat-Bison@002) & Google & No & No & Poe & United States \\
Coral (Command) & Cohere & Yes & No & Cohere & Canada \\
Mistral-7B-Instruct & Mistral & No & Yes & Poe & France \\
Bard (PaLM 2) & Google & Yes & No & Google & United States \\
Falcon 180B & Technology Innovation Institute & No & No & Huggingface & United Arab Emirates \\
Qwen-7B-Chat & Alibaba Cloud & No & Yes & Modelscope & China \\
\bottomrule
\end{tabular}
\label{tab:model_details}
\end{table}
\end{landscape}

\noindent included detailed question background, resolution criteria, and question text as they were posed on the public forecasting tournament, see Figure \ref{fig:Prompt1}.

For every set of machine forecasts, we also recorded the publicly available median human crowd prediction at the end of the day that the machine forecast was entered. If the prediction was entered on the first day, we collected the human crowd predictions at the end of the second day that the question was open to allow for higher participation rates. This was done to ensure a fair comparison of machine and human forecasts, as many LLMs can recall the current date, thus making timed forecasts of the nature studied here potentially sensitive to asynchronous queries while also introducing bias with respect to the human crowd. For roughly half the questions, the human forecasters were not able to see the human crowd forecast, though there is significant heterogeneity when the community predictions were made available to human forecasters. In 15 out of 31 questions, our data was collected prior to the revelation of the community prediction to the human forecasters. 

\begin{figure}[h]
    \centering
    \includegraphics[width=\textwidth]{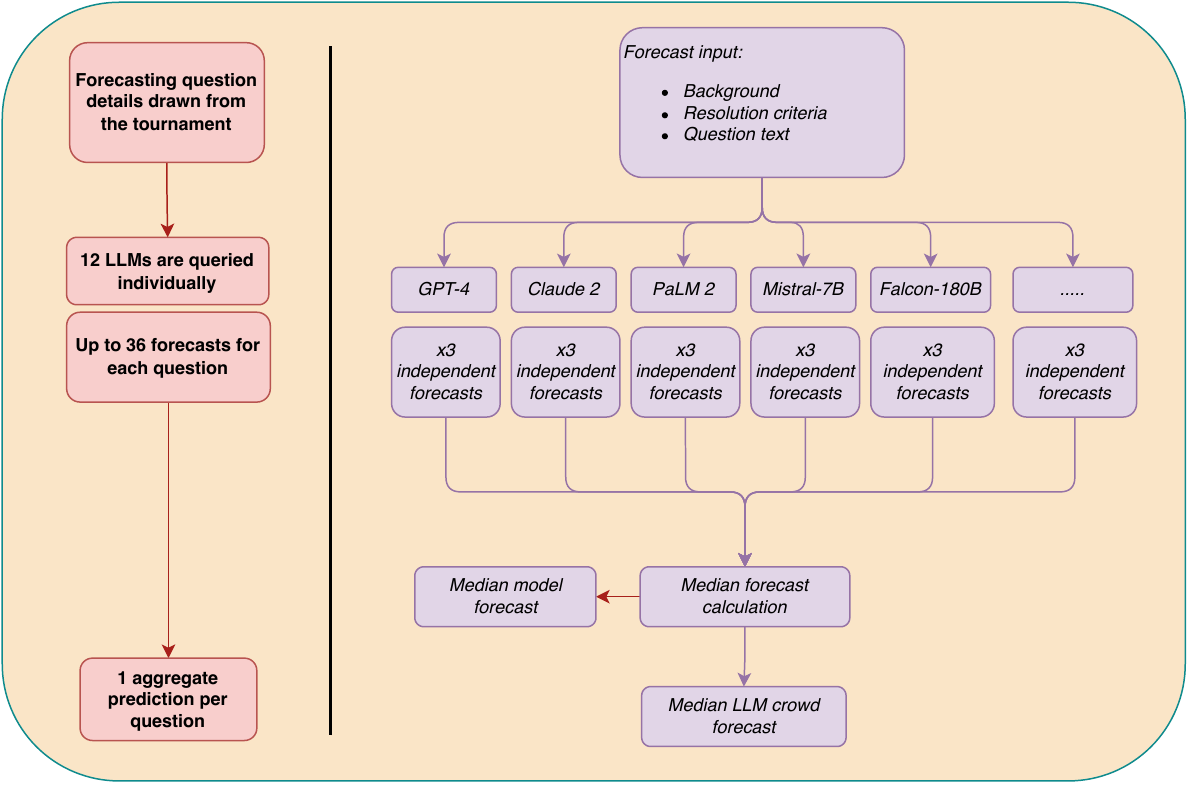} 
    \caption{LLM Ensemble Mechanism Overview}
    \label{fig:figure7}
\end{figure}

For the human forecasts, we took the publicly available median forecast for each question. For the LLM ensemble approach, we computed the median from all non-missing forecasts on each question. We also computed the median forecast on each question for each model to enable cross-model comparisons. See Figure \ref{fig:figure7} for an overview of our LLM ensemble approach.

\subsection{Study 2}

In Study 2, we focused exclusively on two frontier models, GPT-4 and Claude 2. We used the same real-world forecasting tournament as in Study 1 as our study context, functioning as a source of questions and human forecasts. For Study 2, we employed a within-model research design that collected two forecasts (pre- and post-intervention) per model run for each question, with each question being posed three times at the standard temperature settings, resulting in six forecasts per model for each question. Our goal was to investigate LLM updating behaviour with respect to human cognitive output, i.e., whether and how LLMs take into account the human prediction estimates that forecasting tournament aggregates provide. We queried GPT-4 and Claude 2 via the OpenAI and Anthropic websites respectively. 

We used a significantly longer and more elaborate set of prompts than in Study 1. The first prompt built on the ’10 commandments of superforecasting’ \parencite{tetlock2016superforecasting} as well as the literature on forecasting and updating, instructing models to carefully consider distinguishing different degrees of doubt, strike the correct balance between under- and overconfidence, and break difficult problems into sub-problems that are easier to solve, among other instructions. The second prompt, the intervention, informed the model of the respective human crowd’s median forecast and asked it to update, if necessary, and to outline the reasons for the update (if any). For a full text of both prompts, see Figure \ref{fig:Prompt2} and Figure \ref{fig:Prompt3}.

\begin{figure}[h!]
\centering
\begin{tcolorbox}[title=Initial Prompt]
\noindent
In this chat, you are a superforecaster who has a strong track record of accurate forecasting. You evaluate past data and trends carefully for potential clues to future events, while recognising that the past is an imperfect guide to the future so you will need to put probabilities on possible future outcomes (ranging from 0 to 100\%). Your specific goal is to maximize the accuracy of these probability judgments by minimising the Brier scores that your probability judgments receive once future outcomes are known. Brier scores have two key components: calibration (across all questions you answer, the probability estimates you assign to possible future outcomes should correspond as closely as possible to the objective frequency with which outcomes occur) and resolution (across all questions, aim to assign higher probabilities to events that occur than to events that do not occur).

You outline your reasons for each forecast: list the strongest evidence and arguments for making lower or higher estimates and explain how you balance the evidence to make your own forecast. You begin this analytic process by looking for reference or comparison classes of similar events and grounding your initial estimates in base rates of occurrence (how often do events of this sort occur in situations that look like the present one?). You then adjust that initial estimate in response to the latest news and distinctive features of the present situation, recognising the need for flexible adjustments but also the risks of over-adjusting and excessive volatility. Superforecasting requires weighing the risks of opposing errors: e.g., of failing to learn from useful historical patterns vs. over-relying on misleading patterns. In this process of error balancing, you draw on the 10 commandments of superforecasting (Tetlock \& Gardner, 2015) as well as on other peer-reviewed research on superforecasting:

\begin{enumerate}
    \item Triage
    \item Break seemingly intractable problems into tractable sub-problems
    \item Strike the right balance between inside and outside views
    \item Strike the right balance between under- and overreacting to evidence
    \item Look for the clashing causal forces at work in each problem
    \item Strive to distinguish as many degrees of doubt as the problem permits but no more
    \item Strike the right balance between under- and overconfidence, between prudence and decisiveness
    \item Look for the errors behind your mistakes but beware of rearview-mirror hindsight biases
    \item Bring out the best in others and let others bring out the best in you
    \item Master the error-balancing bicycle
\end{enumerate}

Once you have written your reasons, ensure that they directly inform you forecast.

Then, you will provide me with your forecast that is a range between two numbers, each between between 0 and 100 (up to 2 decimal places) that is your best range of prediction of the event. Output your prediction as “My Prediction: Between XX.XX\% and YY.YY\%”. Take a deep breath and work on this problem step-by-step.

The question that you are forecasting as well as some background information and resolution criteria are below. Read them carefully before making your prediction.

\textbf{Background:}

\textbf{Resolution Criteria:}

\textbf{Question:}
\end{tcolorbox}
\caption{Initial prompt for Study 2}
\label{fig:Prompt2}
\end{figure}

\begin{figure}[h!]
\centering
\begin{tcolorbox}[title=Prediction Intervention]
\noindent
You have made your forecast based on careful reasoning and analysis. Now consider the following new piece of information: The median crowd prediction in the forecasting tournament where this question was posed was XXX\%.

Please adjust your reasoning and forecast based on this information, as you deem appropriate. The large research literature on the "wisdom of the crowd" suggests it is difficult for any single forecaster to out-predict crowd medians or averages. But there are occasions when the crowd has proven to be wrong. In considering whether/how much to revise your earlier forecast, keep in mind the theme of error-balancing: the need to balance the risk of giving too little weight to the crowd judgment vs. the risk of over-relying on the crowd. Please explain how you balanced these risks. Please also make this prediction be in the same format as before: “My Prediction: Between XX.XX\% and YY.YY\%”.
\end{tcolorbox}
\caption{Prediction intervention prompt for Study 2}
\label{fig:Prompt3}
\end{figure}

For both prompts, we collected forecasts not as point estimates but as probability ranges between 0\% and 100\% with two decimal point specificity. For further analysis, we treat the midpoint of this range as the point estimate and the provided predictions as upper and lower estimates. The human crowd median that is provided to the models is collected within 48 hours of the community prediction being revealed to allow human forecasters to learn about it and update their forecasts accordingly, generally leading to more well-calibrated predictions. Because of the time difference, the human forecasts are more accurate than those used in Study 1. 

\section{Results}

\subsection{Study 1}

We collected a total of 1007 individual forecasts over the 31 questions from twelve LLMs that make up the ensemble. For 109 forecasts that we did not collect, this was due to technical problems with the model or interface at the time of forecast collection (in the case of Falcon-180B and PaLM 2),or because other models selectively chose not to answer certain questions, presumably due to their content restriction policies (this was the case for Coral (Command) and Qwen-7B-Chat). We also recorded some missing forecasts for Bard, which was due to the fact that the underlying model powering the interface was changed to Gemini Pro. To ensure consistency and allow comparisons between the different contexts of PaLM 2, we stopped collecting data at this point. 

\begin{figure}[h]
    \centering
    \includegraphics[width=0.8\textwidth]{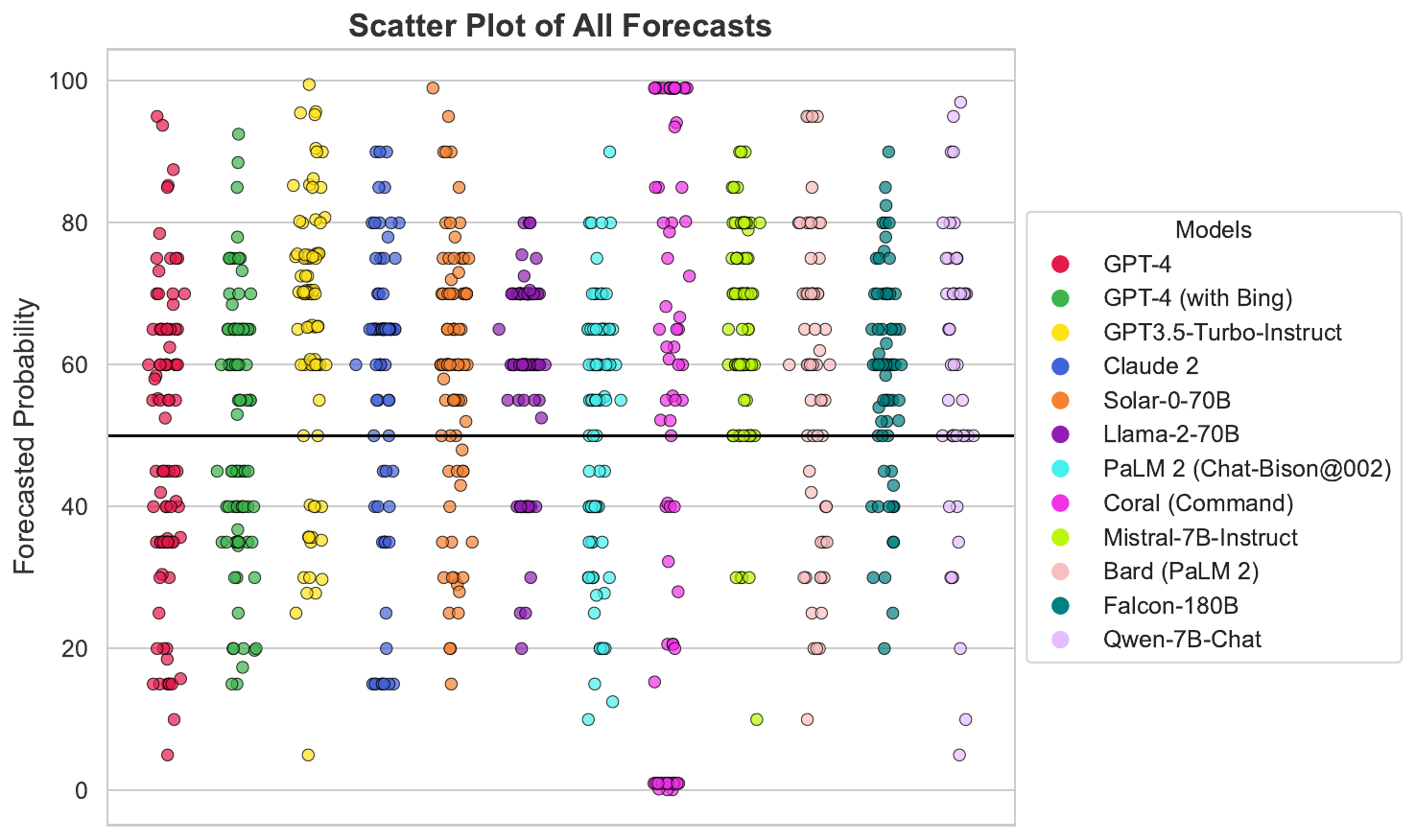} 
    \caption{Scatter Plot of all LLM predictions across all questions}
    \label{fig:figure1}
\end{figure}

 Across all models and questions, we observe a minimum raw forecast value of 0.1\% and a maximum raw forecast value of 99.5\%, with a median forecast of 60\%. This indicates that the LLM models are more likely to make predictions above the 50\% mid-point, with the mean forecast value of the crowd M=57.35 (SD=20.93) being significantly above the 50\% mark, t(1006)=86.20, p<0.001. Importantly, the total question set resolved close to evenly, with 14/31 questions resolving positively. This imbalance thus suggests that LLM predictions generally favour positive resolutions above and beyond the appropriate empirical expectation, with just over 45\% of questions resolving positively. Such a bias towards more positive predictions may be a function of the machine-equivalent of acquiescence bias \parencite{costello2015acquiescence}, where human responders tend to favour the positive/agreement option irrespective of question content \parencite{hinz2007acquiescence}. See Figure \ref{fig:figure1} for a scatter plot of all model forecasts across all questions that shows heterogeneity between models of forecast distribution, ranges, and acquiescence bias.

In order to assess forecasting accuracy, we use the strictly proper scoring rule \parencite{gneiting2007strictly} of Brier scores \parencite{brier1950verification}. It is a metric for assessing the accuracy of probabilistic predictions by taking the mean squared difference between the forecasted probability and the actual outcome. It is defined mathematically as
\[
\text{Brier Score} = (f_i - o_i)^2
\]
where \(f_i\) is the forecasted probability for the instance, and \(o_i\) is the actual outcome, which can be 0 or 1. A lower Brier score indicates higher accuracy, with 0 being the perfect accuracy score. A score of 0.250 represents a typical benchmark that would be arrived at if all predictions were 50\%.

Testing our first hypothesis as preregistered, we investigate whether the LLM crowd can outperform the simple baseline of assigning a 50\% prediction on every question, a baseline that GPT-4 was unable to beat in previous work \parencite{schoenegger2023large}. To arrive at our LLM median forecast for this and further analysis using this aggregate, we calculate the median LLM forecast across all models for every question. We then take these medians and average them across all questions. We then take this average and compare it a Brier score of 0.25 (the result of predicting 50\% on all questions). We are able to reject our null hypothesis, with the LLM crowd, M=0.20 (SD=0.12), being significantly more accurate than the benchmark, t(30) = -2.35, p = 0.026. This is first evidence that crowd-aggregated LLM forecasts can improve upon basic benchmarks. 

Next, we compare the LLM crowd performance to that of the human crowd for our second hypothesis, directly putting the two crowd-aggregation mechanisms head-to-head. To do this, we use the same LLM crowd average as before (taking the median LLM prediction on each question and averaging up the Brier scores across questions). We compare this to the average of median human predictions on the same questions. In our preregistered analysis, we fail to find statistically significant differences between the LLM crowd's mean Brier score of M=0.20 (SD=0.12) and that of the human crowd, M=0.19 (SD=0.19), t(60) = 0.19, p = 0.850. 

\begin{figure}[h]
    \centering
    \includegraphics[width=0.8\textwidth]{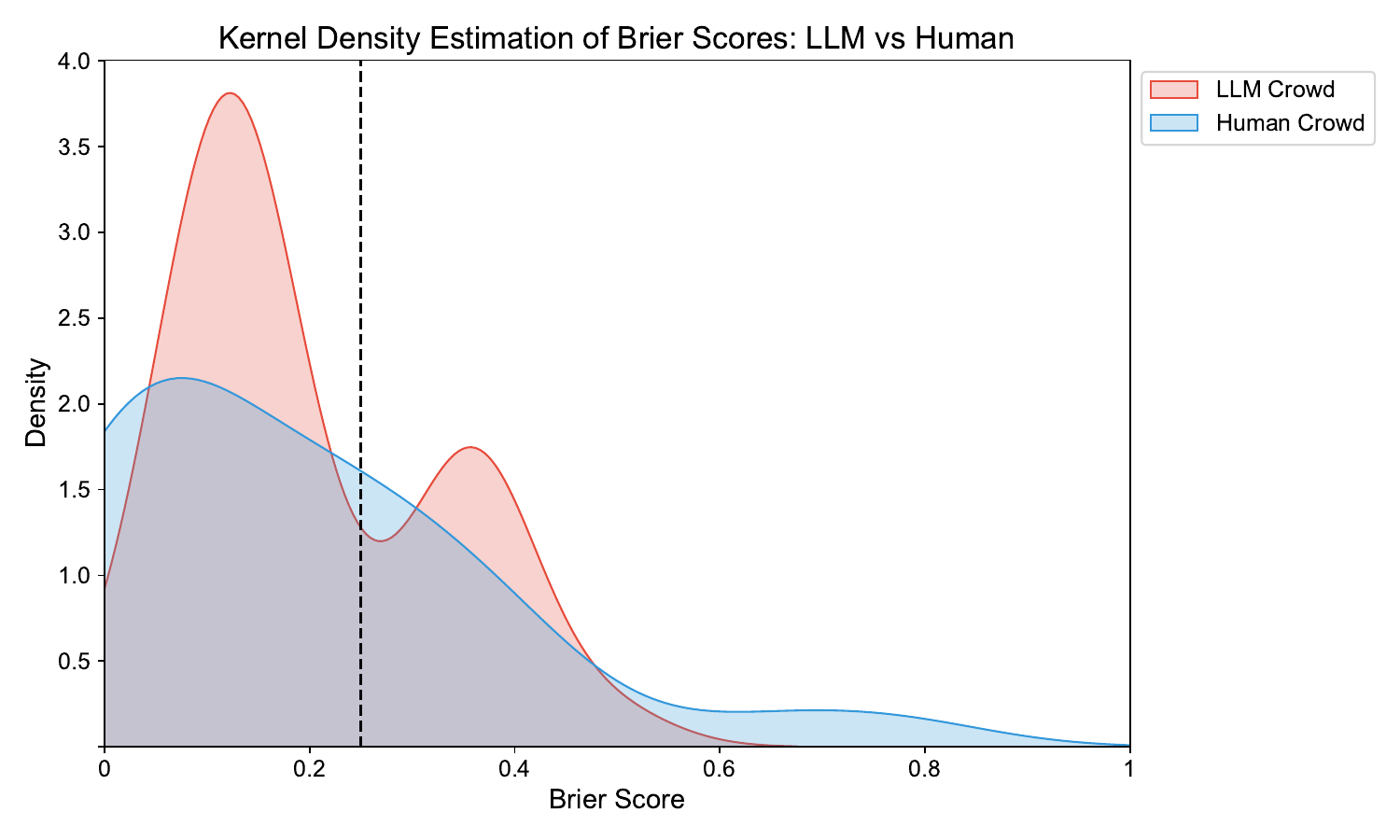} 
    \caption{KDE of the LLM and human crowd forecasts (averaged median scores over all questions). Vertical dotted black line represents the 50\% baseline.}
    \label{fig:figure2}
\end{figure}

This result only enables us to directly conclude that the LLM crowd is neither more nor less accurate than the human crowd in the question set studied here. To provide some evidence in favour of the equivalence of these two approaches, we conduct a non-preregistered equivalence test with the conventional medium effect size of Cohen's d=0.5 as equivalence bounds \parencite{cohen2013statistical}, which allows us to test whether the effect is zero or less than a 0.081 change in Brier scores. For these equivalence bounds, we find that the LLM crowd and the human crowd are equally accurate, with both tests for the lower bound, t(60)=2.16, p=0.017 and the upper bound, t(60)=-1.78, p=0.040, being statistically significant. This provides evidence that the LLM crowd is as accurate as the human crowd within these bounds, though note that the bounds are quite wide.

\begin{table}[h]
\centering
\caption{Average Brier Score for Each Model}
\begin{tabular}{
  l
  S[table-format=1.2]
  S[table-format=1.2]
}
\toprule
{Model} & {Accuracy} & {SD} \\
\midrule
GPT-4 & 0.15 & 0.11 \\
GPT-4 (with Bing) & 0.16 & 0.11 \\
Bard (PaLM 2) & 0.19 & 0.17 \\
Falcon-180B & 0.21 & 0.13 \\
Claude 2 & 0.21 & 0.16 \\
Solar-0-70B & 0.22 & 0.16 \\
PaLM 2 (Chat-Bison@002) & 0.23 & 0.15 \\
Mistral-7B-Instruct & 0.24 & 0.16 \\
Qwen-7B-Chat & 0.24 & 0.17 \\
GPT3.5-Turbo-Instruct & 0.25 & 0.20 \\
Llama-2-70B & 0.25 & 0.16 \\
Coral (Command) & 0.38 & 0.40 \\
\midrule[0.15pt] 
Human                      & 0.19 & 0.19 \\
\bottomrule
\end{tabular}
\label{table:TableBrier}
\end{table}

For our third null hypothesis, we compare the forecasting accuracy of each model (and the human crowd) against each other to find potential effects of internet access (GPT-4 vs. GPT-4 with Bing) or access points (Bard with PaLM2 vs. PaLM2). Using an analysis of variance, we find significant aggregate differences, F(12, 354)=2.64, p=0.002, leading us to reject our third null hypothesis. Using Tukey HSD to adjust for multiple comparisons in the post-hoc pair-wise tests, we find that Coral (Command) underperforms a set of models (e.g., Claude 2, GPT-4) as well as the human crowd. However, we fail to find statistically significant effects between any other pairs not involving Coral (Command), thus being unable to provide evidence in favour or against potential effects of internet access, access points, or fine-tuning on prediction accuracy. See Table \ref{table:TableBrier} for average Brier scores for each model. 

\begin{figure}[h]
    \centering
    \includegraphics[width=0.8\textwidth]{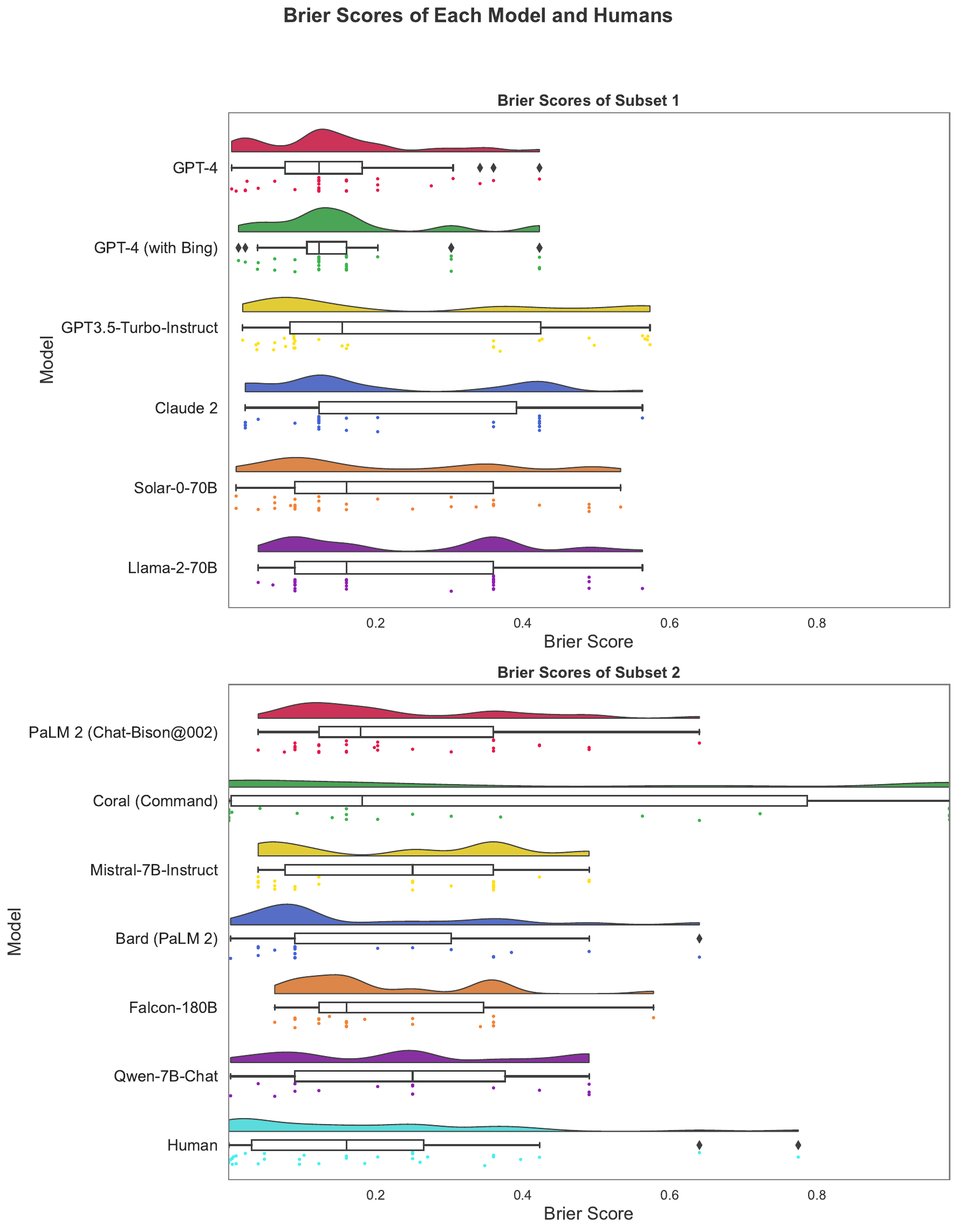} 
    \caption{Raincloud plots of Brier scores for each LLM model as well as the human crowd.}
    \label{fig:figure4}
\end{figure}

For all three hypotheses, we implemented the Benjamini-Hochberg (BH) procedure to adjust the p-values obtained from multiple hypothesis tests. This method was selected to control the False Discovery Rate (FDR) and thereby reduce the risk of Type I errors. The original p-values for null hypotheses 1, 2, and 3 were 0.026, 0.850, and 0.002, respectively. These p-values were first sorted in ascending order and then ranked accordingly. The adjusted p-values were computed using the Benjamini-Hochberg procedure, which calculates the adjusted p-value for the $i$-th hypothesis as $\min\left\{1, \frac{p_i \times m}{i}\right\}$, where $p_i$ is the $i$-th p-value in the sorted list, $m$ is the total number of hypotheses tested, and $i$ is the rank of the p-value. The results show that the adjusted p-values for the hypotheses were 0.039 for the first hypothesis (original p=0.026), 0.850 for the second hypothesis (original p=0.850), and 0.006 for the third hypothesis (original p=0.002). These results indicate that our rejections of the first and third null hypothesis remain robust after adjusting for multiple comparisons.

\begin{figure}[h]
    \centering
    \includegraphics[width=0.8\textwidth]{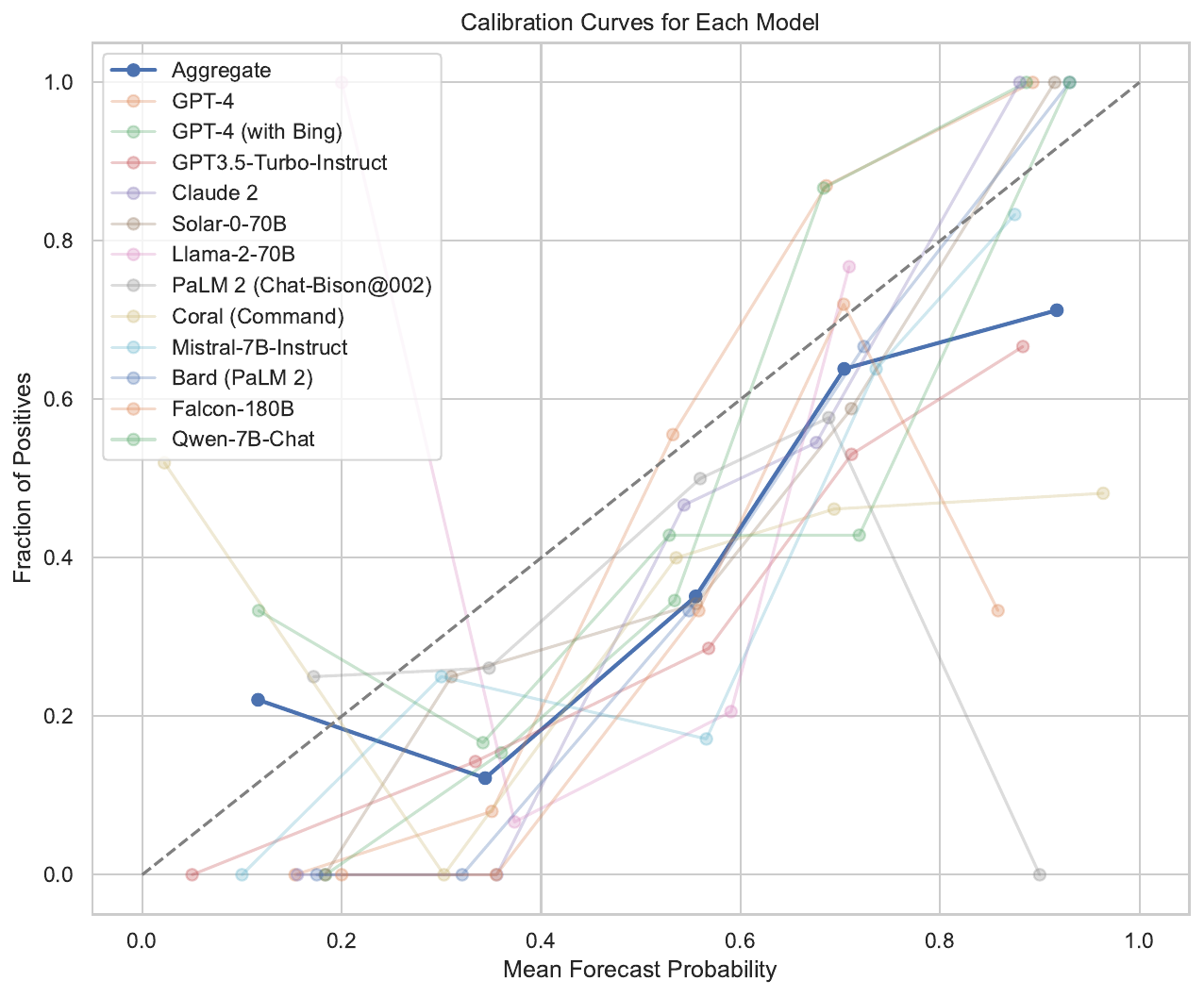} 
    \caption{Calibration plot for all LLM models as well as the aggregate (bolded)}
    \label{fig:figure5}
\end{figure}

In non-preregistered analyses, we conduct calibration analyses using the Murphy Decomposition \parencite{mandel2014accuracy, siegert2017simplifying} to provide data on how well calibrated the LLM models are in this context, i.e., how reliably their probability estimates match the fraction of real outcomes. In Figure~\ref{fig:figure5}, calibration curves for each model and their aggregate are plotted against the ideal 45-degree dotted line. This dotted line represents perfect calibration, where predicted probabilities match observed frequencies. Deviations from this line indicate calibration errors: curves above the line suggest underconfidence (predicting events as less likely than they actually are), while those below indicate overconfidence (predicting events as more likely than they actually are). Figure~\ref{fig:figure5} visually represents how closely the models' predictions align with actual outcomes. We also calculate the Calibration Index (CI), which quantifies this deviation, with lower values indicating better calibration. CI is calculated using the formula:

\[
CI = \frac{1}{N} \sum_{k=1}^{K} N_k (f_k - o_k)^2
\]

where \( N \) is the total number of forecasts, \( K \) the number of bins, \( N_k \) the number of forecasts in bin \( k \), \( f_k \) the mean forecast probability in bin \( k \), and \( o_k \) the observed relative frequency in bin \( k \). This weights each bin's contribution to the Calibration Index (CI) by the number of forecasts it contains. This approach ensures that bins with more forecasts, which provide a more statistically reliable estimate of forecasting accuracy, have a proportionately greater impact on the overall CI.

Our results demonstrate poor calibration of most models and overconfidence of the aggregate, suggesting that models overpredict outcomes compared to their actual rate of occurrence, see Figure \ref{fig:figure5}. This is in line with the finding that we find a acquiescence bias of LLMs on a question set where less than half of questions resolve positively. We also find generally poor calibration across all models. However, there are substantial differences in the CI scores, with some models having substantially better calibration than others, see Table \ref{table:TableCI}. This suggests that a further line of research may build upon improving calibration of models in an attempt to improve machine prediction capabilities and reliability further.

\begin{table}[h!]
\caption{Calibration index values for all LLM models.}
\label{table:TableCI}
\centering
\begin{tabular}{
  l
  S[table-format=1.2, round-mode=places, round-precision=3]
}
\toprule
\textbf{Model} & \textbf{Calibration Index} \\
\midrule
Falcon-180B                 & 0.027 \\
Qwen-7B-Chat                & 0.055 \\
PaLM 2 (Chat-Bison@002)     & 0.068 \\
Bard (PaLM 2)               & 0.071 \\
Llama-2-70B                 & 0.071 \\
GPT-4                       & 0.075 \\
Mistral-7B-Instruct         & 0.080 \\
Solar-0-70B                 & 0.081 \\
Claude 2                    & 0.082 \\
GPT-4 (with Bing)           & 0.088 \\
GPT3.5-Turbo-Instruct       & 0.106 \\
Coral (Command)             & 0.212 \\
\midrule[0.15pt]
Aggregate                   & 0.041 \\
\bottomrule
\end{tabular}
\end{table}

\subsection{Study 2}

For Study 2, we collected a total of 186 primary forecasts and 186 updated forecasts from both frontier models (GPT-4 and Claude 2) over the 31 binary questions studied. Neither model refused to provide a forecast or failed to respond to our querying. 

First, we test whether exposure to the human crowd median improves model accuracy. We are able to reject the first null hypothesis of Study 2 for both models: For GPT-4, there is a statistically significant difference in Brier Scores before and after exposure to the human median, with an average Brier score for the primary forecast of 0.17 (SD: 0.13) and an updated score of 0.14 (SD: 0.11), p = 0.003. For Claude 2, we also find a statistically significant difference in Brier Scores before and after exposure to the human median, improving from 0.22 (SD: 0.19) to 0.15 (SD: 0.14), p < 0.001. This suggests that the provision of human cognition in the form of crowd forecasts can improve model prediction capabilities. 

We also find that, testing our second hypothesis, the size of the prediction interval narrows after exposure to human crowd predictions that lie within the probability range provided by the model, as would be predicted by theory: The prediction intervals for GPT-4 become significantly narrower after exposure to the human median, ranging from an average interval size of 17.75 (SD: 5.66)to 14.22 (SD: 5.97), p < 0.001. The prediction intervals for Claude 2 also become significantly narrower after exposure to the human median forecast, narrowing from 11.67 (SD: 4.201 to 8.28 (SD: 3.63), p <0.001. This suggests that the models appropriately reduce their prediction uncertainty when the human forecast is already included in the LLM's, incorporating this additional information and adjusting their uncertainty. See Figure \ref{fig:figure4} for a graphical illustration of LLM forecasts for either model before and after exposure to the human forecasts.

\begin{figure}[h]
    \centering
    \includegraphics[width=0.8\textwidth]{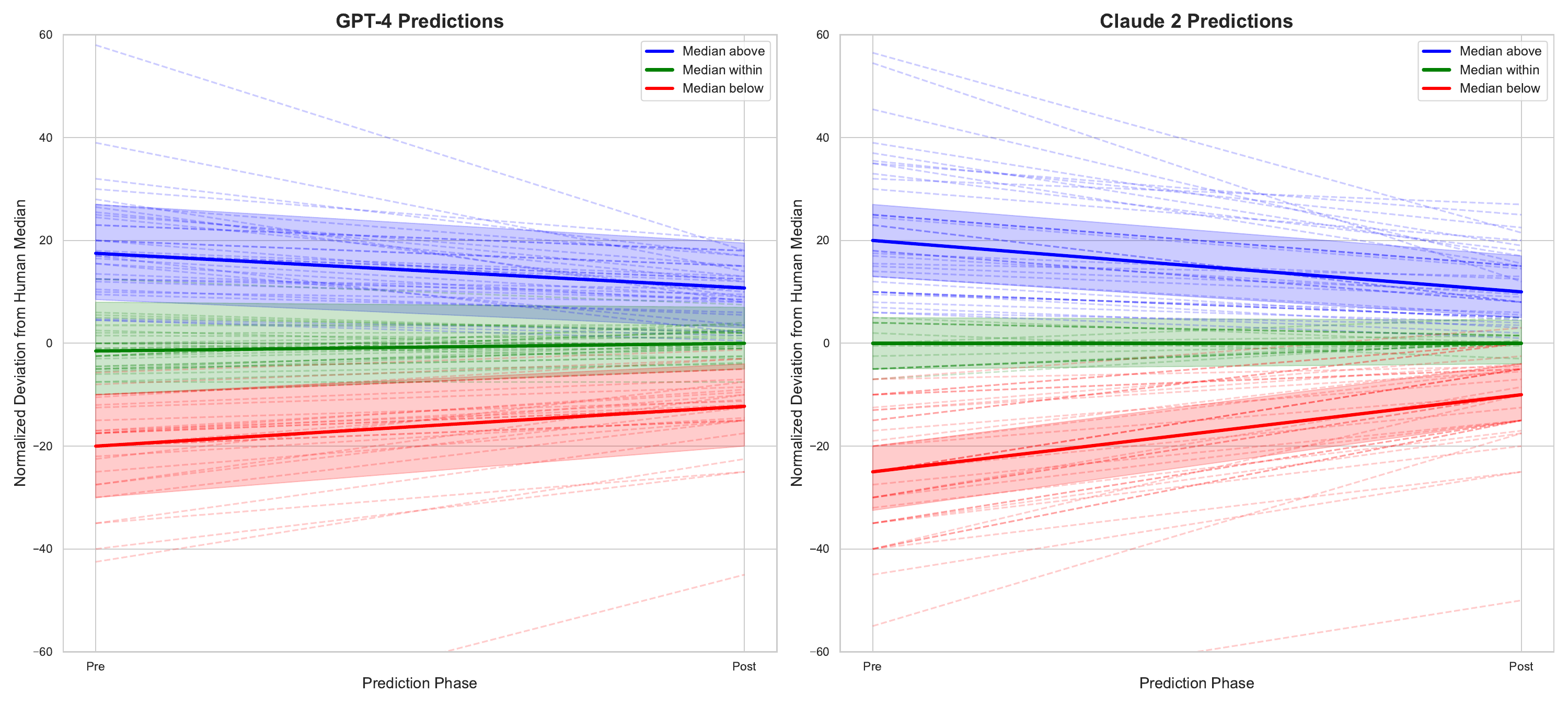} 
    \caption{LLM forecasts for GPT-4 (left) and Claude 2 (right) before and after exposure to the human forecast. Colours distinguish first forecasts above, below, or within 20 percentage points of the human median forecast. Highlighted changes and intervals are of the respective median forecast within that group.}
    \label{fig:figure4}
\end{figure}

Lastly, with respect to our third hypothesis, we analyse whether LLMs' updates are proportional to the distance between their point forecast and that of the human benchmark. We are able to reject our null hypothesis for both models, finding significant correlation between the initial deviation and the magnitude of forecast adjustment for GPT-4, r=0.88, p < 0.001 as well as for Claude 2 r=0.87, p < 0.001. This suggests that models move their predictions roughly in accordance with how large the difference between their prediction and the human median is.

As in Study 1, we use the Benjamini-Hochberg procedure for controlling multiple comparisons, given our three hypotheses each tested for each model, resulting in six tests. The original p-values were [0.001, 0.001, 0.001, 0.001, 0.001, 0.003]. After applying the Benjamini-Hochberg adjustment, the p-values were [0.006, 0.006, 0.006, 0.006, 0.006, 0.003], all of which were below the 0.05 FDR threshold. This indicates that, post-adjustment, the results from all tests remained statistically significant.

We also conduct the following exploratory analysis. Instead of comparing the LLM forecast after having been exposed to the human median to the LLM forecast before this exposure as preregistered, we compare this updated prediction to a simple average of the machine and human predictions as a naive benchmark using straightforward aggregation. This allows us to test whether the improvements the models make are due to understanding of the need to appropriately update or simply as an agreement-focused response. We find in paired t-tests that for both GPT-4 at a Brier score of 0.13, \( t(92) = 2.583, p = .011 \), and Claude 2 at a Brier score of 0.14, \( t(92) = 3.530, p = .001 \), their updated forecasts are significantly less accurate than a simple average between the machine and the human median forecasts. This suggests that the updating itself is directionally correct but fails to improve upon a simple benchmark.

\section{Discussion}

Our results show that LLM prediction capabilities can rival the gold standard of the human crowd tournament method, if they themselves draw on what we call the `wisdom of the silicon crowd.' Previous results on single models \parencite{schoenegger2023large,halawi2024approaching} showed that LLMs not only underperformed compared to a human crowd in a probabilistic forecasting context, but also failed to clear simple benchmarks; while others \parencite{abolghasemi2023humans} failed to find evidence in favour of the LLMs outperforming humans in the context of time-series forecasting.\footnote{For more applications of LLMs in time-series forecasting see additional work \parencite{cholakov2021transformers, jin2023time, gruver2024large}} However, taking into account more sophisticated systems built on top of LLMs, such as combined retrieval and reasoning systems \parencite{halawi2024approaching}, human-level prediction accuracy may already be considered matched in some aspects. We propose that the capabilities jump in moving from single frontier models to crowds of simple models in the same probabilistic forecasting context is a benefit that can be exploited in a variety of real-world contexts, as this aggregation approach remains simple to implement and does not require additions like that of news retrieval on each question. Our finding opens the door for simple, practically applicable steps like forecast aggregation to increase current AI models' forecasting ability---to predict future events in politics, economics, technology, and other real-world subjects---to a level on par with the human crowd. This opens up a lot of directly applied work, given that LLM prediction capabilities can inform decision-makers and businesses in circumstances where accurate probabilistic forecasts are difficult or expensive to acquire. Furthermore, since both our finding and the finding of \textcite{halawi2024approaching} suggest that placing individual LLMs in advanced systems can increase their forecasting ability to a market-competitive level, it is natural to expect LLM predictions to be more widely applied across society in the near future.

Importantly, our finding holds despite the presence of an acquiescence bias \parencite{costello2015acquiescence,hinz2007acquiescence} in model predictions, in that our models' predictions are more likely to be above 50\%, despite the resolution rate of all questions being almost even. This suggests that the `wisdom of crowds' effect using median as our aggregation is able to counteract even this acquiescence bias that is present in the majority of individual models, a robustness feature of the `wisdom of crowds' mechanism \parencite{koriat2012two}. In our aggregation results, we also find that only three of the twelve models outperform the model median, which is also in line with standard accounts of wisdom of crowds. This overall suggests that the `wisdom of the crowd' effect---in addition to applying to the human context---also applies to the silicon context. The literature on the size of the crowd needed to produce reliable `wisdom of the crowd' effects is not very well established, though a central finding is that increasing crowd size does lead to better performance \parencite{walter2022measuring}. As such, a natural next step in this line of research is to expand the set of models queried from the twelve we used to a substantially higher number. 

However, there also do remain substantial areas of improvement for machine predictions in probabilistic forecasting. Most directly, both the aggregate and the individual models were badly calibrated, with most models showing overconfidence, i.e., they assign higher probabilities to outcomes than is warranted by the empirical facts. Improving calibration is central to providing reliable predictions over the long run \parencite{buizza2018ensemble}, and our results of acquiescence bias suggest that this may be an actionable area for future work. Additionally, as the distance between the end of the training data and the forecasted period grows, forecasts may become less accurate as necessary background knowledge is no longer readily available to the model. Moreover, our study could draw on a well-curated set of questions from the forecasting tournament. Applications in contexts where neutral background information and question details are not available may further reduce performance. 

Our results from Study 2 contribute to the literature on human-AI interactions \parencite{yang2024human, kim2024human}. While previous work in the context of forecasting has looked at how LLMs can augment humans in improving prediction accuracy \parencite{schoenegger2024ai}, this paper provides evidence for the reverse. Specifically, our results show that machine predictions can be improved substantially by the provision of human cognition output drawn from forecasting tournaments. This finding suggests at first glance that LLM reasoning is already advanced enough to properly exploit the informational value provided by human cognition output. However, our exploratory analyses find that this process is substantially less effective than simply averaging the two estimates, suggesting that aggregation methods based on the reasoning capabilities of frontier models (in this case, GPT-4 and Claude 2) still underperform simple aggregations.
    
On the other hand, our findings that both frontier models (GPT-4 and Claude 2) respond as expected in their forecast updates---reducing their uncertainty when the human estimate lies within their prediction intervals, and updating in relation to the distance between their own point estimate and the human forecasts---match past theory and results pertaining to human forecasters \parencite{atanasov2020small}. This overall suggests that the ability of these models to reason and act as expected---by past theory and results pertaining to human forecasters---depend on the type of task and benchmark applied. While this is not a massively strict test of their reasoning abilities---as alternative explanations of model behaviour being explained by simple expectation fulfilling remain---it does provide some evidence in favour of it. 

Importantly, both studies reported in this paper test LLM capabilities in a context where it is not possible that any of the answers used to resolve the questions were part of the training data, as we queried the models in real-time alongside the human tournament. Because all question answers were unknown at the time of data collection---even to the study authors---this provides an ideal evaluation criterion for LLM capabilities: one at which our LLM ensemble approach excelled. Our findings provide evidence of LLMs' advanced reasoning capabilities, and does so in a robust way such that many of the challenges that may be raised with respect to traditional benchmarks do not apply.

In conclusion, the present paper is among the first to show that current LLMs are able to provide a human-crowd-competitive level of accurate forecasting about future real-world events. In order to do so, it is sufficient to apply the simple, practically applicable method of forecast aggregation: manifesting as the LLM ensemble approach in the so-called silicon setting. This replicates the human forecasting tournament's `wisdom of the crowd' effect for LLMs: a phenomenon we call the `wisdom of the silicon crowd.' Our finding opens up a number of areas for further research as well as practical applications, since the LLM ensemble approach is substantially cheaper and faster than data collection from human forecasters. Future research can aim to combine the ensemble approach with model and scaffolding progress, which may potentially result in even stronger capability gains in the domain of forecasting. 

\section*{Acknowledgements}

We are grateful to Lawrence Phillips and Peter M\"{u}helbacher for helping us discover and correct a coding error in the non-preregistered equivalence test pertaining to the second null hypothesis of Study 1.

\printbibliography

\newpage

\begin{longtable}{p{\textwidth}}
\caption{Full list of questions}\label{tab:full_list_questions} \\
\toprule
\textbf{Questions} \\
\midrule
\endfirsthead

\multicolumn{1}{c}%
{{\bfseries \tablename\ \thetable{} -- continued from previous page}} \\
\toprule
\textbf{Questions} \\
\midrule
\endhead

\midrule
\multicolumn{1}{r}{{Continued on next page}} \\*
\endfoot

\endlastfoot

Will a nearly continuous human chain stretch across the length of the Forth and Clyde Canal on 14 October 2023? \\
\midrule
Will Hamas lose control of Gaza before 2024? \\
\midrule
Will Yahya Sinwar cease to act as Hamas Chief in the Gaza Strip before 2024? \\
\midrule
Will Israel carry out and explicitly acknowledge a deadly attack on Iran before 2024? \\
\midrule
Will the Conservatives hold on to their seat in the Mid Bedfordshire by-election? \\
\midrule
Will it be determined that Israel was responsible for the attack on the Al-Ahli Baptist Hospital in Gaza City before 2024? \\
\midrule
Will the Federal Funds Rate be raised before December 14, 2023? \\
\midrule
Will Peter Bone MP be suspended from Parliament in 2023? \\
\midrule
Will George Weah win re-election in the 2023 Liberian General Election? \\
\midrule
Will India request that another Canadian diplomat be recalled before 2024? \\
\midrule
Will New Delhi experience a "Very Unhealthy" or worse air quality index on at least four of the seven days for the week starting October 29? \\
\midrule
Will Mike Johnson remain Speaker until 2024? \\
\midrule
Will there be an additional Russian IPO on the MICEX in 2023? \\
\midrule
Will Donald Trump spend at least one hour confined in a jail cell before January 1, 2024? \\
\midrule
Will the second Starship integrated flight test achieve liftoff before January 1, 2024? \\
\midrule
Will Sarah Bernstein or Chetna Maroo win the 2023 Booker Prize? \\
\midrule
Will Bitcoin reach \$40,000 before January 1, 2024? \\
\midrule
Will Volodymyr Zelenskyy visit Israel before 2024? \\
\midrule
Will Delhi perform cloud seeding before December 1, 2023? \\
\midrule
Will the MONUSCO UN peacekeeping mission to the Democratic Republic of the Congo be extended with a military personnel ceiling above 11,000 before January 1, 2024? \\
\midrule
Will OpenAI report having $\ge$99\% uptime for ChatGPT and the OpenAI API in December 2023? \\
\midrule
Will the November 2023 Israel-Hamas humanitarian pause be extended? \\
\midrule
Will a majority of voters approve Venezuela's referendum on incorporating Guayana Esequiba into Venezuela? \\
\midrule
Will any additional Republican candidates for president drop out before 2024? \\
\midrule
Will there be a white Christmas in at least 4 of these 9 large European cities in 2023? \\
\midrule
Will the US Supreme Court issue a decision on hearing the case about presidential immunity before January 1, 2024? \\
\midrule
Before 2024, will it be announced that either of the Harvard or MIT presidents will vacate their positions? \\
\midrule
Will a major shipping company announce that they are resuming shipments through the Red Sea before 2024? \\
\midrule
Will the ban on imports of Apple watches with blood oxygen sensors take effect before December 27, 2023? \\
\midrule
Will there be a US military combat death in the Red Sea before 2024? \\
\midrule
Will NASA re-establish communications with Voyager 1 before 1 Jan 2024? \\
\bottomrule
\end{longtable}

\end{document}